\documentclass[aps,pre,twocolumn,floatfix,showpacs,superscriptaddress]{revtex4-2} 
\usepackage{amssymb,amsmath}
\usepackage{graphicx}
\usepackage{subfigure}
\usepackage[english]{babel}
\usepackage{float}
\usepackage{color}
\usepackage{xcolor}
\usepackage{comment}
\usepackage{mathtools} 
\usepackage{xspace} 
\usepackage[hidelinks]{hyperref}

\usepackage{comment}

\newcommand{\ve}[1]{\mathbf{#1}} 

\renewcommand{\exp}[1]{\mathchoice{\mathrm{e}^{#1}}{\operatorname{exp}\left(#1\right)}{\operatorname{exp}\left(#1\right)}{\operatorname{exp}\left(#1\right)}}
\newcommand{\ExpNB}[1]{\operatorname{exp}}

\newcommand{\ave}[1]{\mathchoice{\left\langle #1 \right\rangle}{\langle #1 \rangle}{\langle #1 \rangle}{\langle #1 \rangle}}

\newcommand{\elabel}[1]{\label{eqn:#1}}
\newcommand{\eref}[1]{(\ref{eqn:#1})}

\newcommand{\flabel}[1]{\label{fig:#1}}
\newcommand{\fre}[1]{~\ref{fig:#1}}
\newcommand{\fref}[1]{Fig.~\ref{fig:#1}}
\newcommand{\frefs}[1]{Figs.~\ref{fig:#1}}

\newcommand{\latin}[1]{{\it #1}}
\newcommand{\ie}{\latin{i.e.}\@\xspace}

\newcommand{\demand}{\stackrel{!}{=}}
\usepackage{xcolor}

\begin{document}

\title{Eph-ephrin-mediated differential persistence as a mechanism for cell sorting}

\author{Marius Bothe}
\email{m.bothe19@imperial.ac.uk}
\address{Department of Mathematics, Imperial College London, 180 Queen’s Gate, London SW7 2BZ, United Kingdom}
\author{Eloise Lardet}
\address{Department of Mathematics, Imperial College London, 180 Queen’s Gate, London SW7 2BZ, United Kingdom}
\author{Alexei Poliakov}
\address{Locomizer Ltd, London, United Kingdom}
\author{Gunnar Pruessner}
\address{Department of Mathematics, Imperial College London, 180 Queen’s Gate, London SW7 2BZ, United Kingdom}
\author{Thibault Bertrand}
\address{Department of Mathematics, Imperial College London, 180 Queen’s Gate, London SW7 2BZ, United Kingdom}
\author{Ignacio Bordeu}
\email{ibordeu@uchile.cl}
\address{Departamento de F\'isica, Facultad de Ciencias F\'isicas y Matem\'aticas, Universidad de Chile, Chile}

\begin{abstract}
The phenomenon of cell sorting/segregation, by which cells organise spatially into clusters of specific cell type or function, is essential for tissue morphogenesis. This self-organization process involves an interplay between mechanical, biochemical, and cellular mechanisms that act across various spatial and temporal scales. Several mechanisms for cell sorting have been proposed; however, the physical nature of these mechanisms and how they lead to symmetric or asymmetric cell sorting remains unclear. Here, using experimental data from cocultures of genetically modified Human Embryonic Kidney (HEK293) cells and numerical simulations, we show the existence of a cell sorting mechanism based on transient increases in the persistence of motion of cells. This mechanism is activated on cells overexpressing the ephrinB1-related receptor EphB2 after their interaction with cells overexpressing ephrinB1. We show that this mechanism is sufficient to cause cell sorting, breaking the symmetry of the sorting dynamics, and show that the duration of this \textit{transient differential persistence} state is optimal for enhancing sorting. Furthermore, we show that in combination with other interaction mechanisms, such as changes in direction—also known as contact inhibition of locomotion—and adhesion forces, differential persistence significantly reduces the timescale of sorting. Our findings offer insights into the behaviour of cell mixtures, the relevance of non-reciprocal interactions, and may provide insight into developmental processes and tissue patterning. 
\end{abstract}

\maketitle

\section{Introduction}

During embryonic development, a series of collective cellular processes take place, driving collective cellular migration, tissue patterning, and specification \cite{mayor2016front,chan2017coordination}.
These processes are influenced by chemical, mechanical, or geometric environmental cues that provide spatial modulation of cell-cell interactions \cite{heller2015tissue}. Non-reciprocal interaction mechanisms, such as contact-inhibition of locomotion (CIL) \cite{abercrombie1954observations,davis2015inter}, can lead to spatial sorting of distinct cell populations, which plays a crucial role in various tissue patterning processes during embryonic development, tissue regeneration and cancer \cite{wilson1907some,astin2010competition,stramerMechanismsVivoFunctions2017}.

Some of the best-studied mechanisms of cell sorting include: differential adhesion \cite{townes1955directed,steinberg1963reconstruction,pawlizakTestingDifferentialAdhesion2015,taylorCellSegregationBorder2017,beatrici2017mean}, differential cortical tensions \cite{ONeill2016,Kindberg2021,mehes20233d} and CIL \cite{abercrombie1954observations,stramerMechanismsVivoFunctions2017}. In general, all these mechanisms involve differential changes in the direction of motion upon contact between cells of the same type (homotypic) or different types (heterotypic). Differential adhesion occurs when cells form stronger adhesive bonds with cells of a preferred type due to specific affinity between membrane-bound adhesion complexes. Differential cortical tension proposes that differences in actomyosin-controlled cell surface tension influences homotypic and heterotypic cell-cell interactions, resulting in a differential adhesion-type dynamics \cite{lucia2022cell}. Both of these mechanisms refer to differences in membrane properties. On the other hand, CIL relies on the polarisation of protein complexes within the cell upon homotypic or heterotypic cell-cell contacts, resulting in a change in the direction of motion of one or both of the cells involved \cite{stramerMechanismsVivoFunctions2017}.

\begin{figure}[t]
\begin{centering}
    \includegraphics[width=0.5\textwidth]{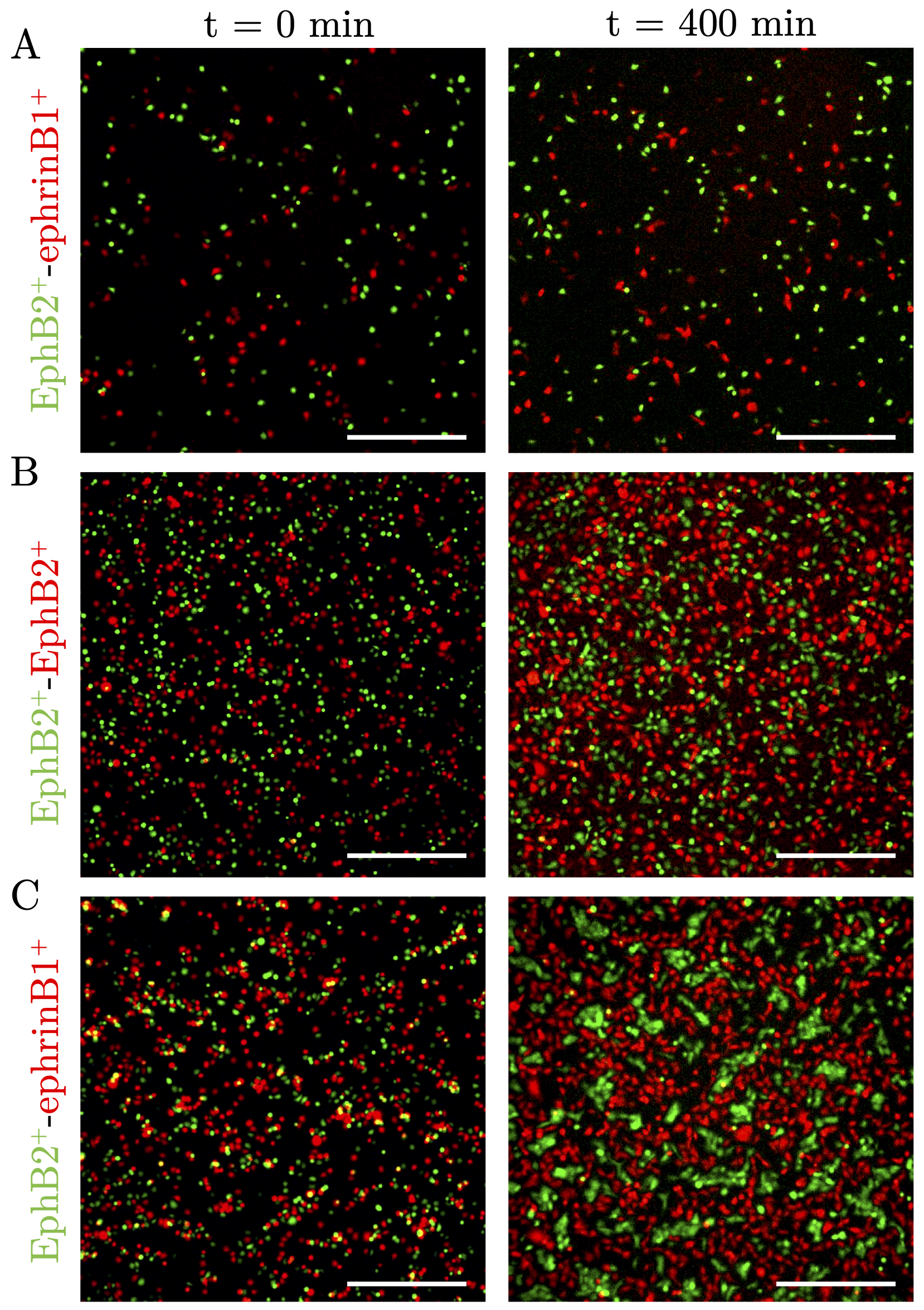} 
    \caption{Snapshots of the initial ($t=0$ min, left) and final ($t=400$ min, right) configurations for different cell mixtures and densities: A) low density of distinct EphB2 (green) and ephrinB1 (red) cell populations; B) No-sorting (control) condition, high densities of same-type EphB2 (green) and EphB2 (red) cells; C) Sorting condition, high density of distinct EphB2 (green) and ephrinB1 (red) cell populations. Scale bar 500 $\mu$m.}
    \flabel{fig:experiments} 
\end{centering}
\end{figure}

Despite the large body of work dedicated to the study of cell sorting, a number of fundamental questions remain open. The mechanisms described above are essentially pair interaction mechanisms, which would in principle work well in dilute scenarios. Whether CIL is important at all at confluence or close to it is still debated. Moreover, these mechanisms refer mainly to changes in \textit{direction} of motion caused by a shift in the cell polarity after interacting with another cell. These changes in cell polarisation may also temporarily affect the persistence of motion of the cells \cite{yolland2019persistent,vaidvziulyte2022persistent}, which in some systems is correlated to movement speed \cite{maiuri2015actin}. In addition, a previous numerical study showed that changes in persistence can cause demixing in a mixture of two types of particle types, provided a quorum sensing-type mechanism by which particles increase their persistence when surrounded by particles of a different type \cite{strandkvistKineticMechanismCell2014}. However, little is known about how such an increase in orientational memory may control cell sorting in real cellular systems, and how this mechanism may interact with other cell sorting mechanisms, such as CIL and differential adhesion. Thus, we questioned whether transient change of persistence of one cell type, activated upon heterotypic interactions (i.e. with cells of a different type) may drive cell sorting in cell mixtures. 

To answer this question, here we propose a simple experimental setup in which two populations of genetically modified human embryonic kidney (HEK293) cells interact non-reciprocally, with one causing a transient change in the persistence of the other. The cell lines used correspond to \cite{poliakov2008}: an ephrinB1$^+$ line, which over-expresses the membrane ligand ephrinB1, and an EphB2$^+$ line, over-expressing the ephrinB1-related receptor EphB2. The Eph-ephrin signalling pathway has diverse functions during embryonic development, including the regulation of cell migration, guidance and sorting, for example during neurogenesis \cite{poliakov2004diverse,jorgensen2009cell, stramerMechanismsVivoFunctions2017,wilkinson2021interplay,bush2022cellular}. Earlier works have shown that cocultures of these and other cell lines exhibit relatively fast cell sorting \cite{poliakov2008,mehes2012collective}, and that Eph-ephrin signalling drives arterial specification and cell sorting \textit{in-vivo} \cite{stewen2024eph}. Coculturing both cell lines provides the means to study differential cell-cell interactions and their effect in cell sorting \cite{poliakov2008,taylorCellSegregationBorder2017} (see  \fref{fig:experiments}).

Through quantitative data analysis, we show that heterotypic interactions result in a differential change in persistence in the motion of one of the interacting cells, which drives cell sorting, without affecting cell speed. Furthermore, we propose a minimal theoretical model and study the effects of transient persistence changes in the dynamics and time-scale of cell sorting. We show how the differential persistence breaks the symmetry of the cell sorting dynamics, and that it may significantly accelerate the sorting dynamics when acting in combination with other cell sorting mechanisms. Depending on the model parameters, cell sorting results in distinct spatial patterns, including mixed phases, phase and micro-phase separation, and regular spatial patterns. Finally, we discuss the implications of our findings for biological systems. 

\subsection*{Experimental setup}

We considered three experimental conditions (see Appendix~\ref{coculture} for details): A low-density coculture of EphB2$^+$ (green) and ephrinB1$^+$ (red) cells (\fref{fig:experiments}A), and two high-density cocultures: a non-segregating (control) coculture of two groups of functionally identical EphB2$^+$ cells, one with green dye CMFDA and the other with orange dye CMRA, which appears red in the figures (\fref{fig:experiments}B), and a segregating coculture of EphB2$^+$ (green, CMFDA dye) and ephrinB1$^+$ (red, CMRA dye) cells (\fref{fig:experiments}C). 
For each condition, cells were imaged every 2 minutes for 400 minutes (approximately 7 hours).
In our experiments, cells have a typical cell diameter of 9 pixels or 30 $\mu$m. Thus, in the following analysis we consider that two cells whose 2d-projected centres are closer than 30 $\mu$m are effectively \textit{in contact} with each other. 

\begin{figure*}[ht!]
\begin{centering}
    \includegraphics[width=1\textwidth]{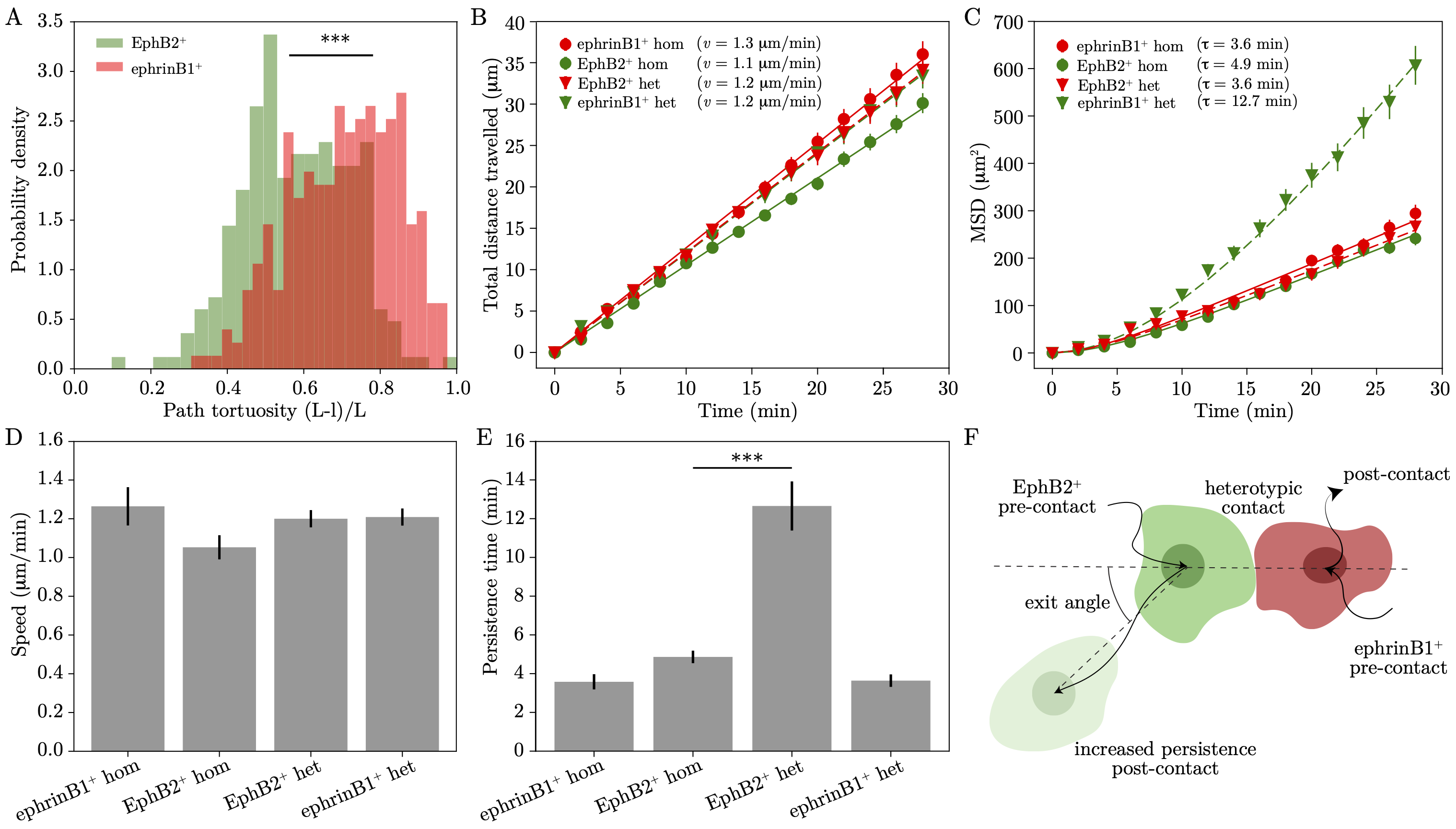} 
    \caption{Motile properties after cell-cell interactions, showing: A) Histogram of tortuosities in the low-density experiment ($***$ indicates $p=9.5e^{-14}$, KS test), B) the total distance travelled and C) mean squared displacement of different cell types after heterotypic (het) and homotypic (hom) contacts (mean and SD) and fits of Eq.~\ref{eq:msd}). D) Estimated self-propulsion speed $v$ for linear fits in (B), and E) persistence time $\tau$ from fits in (c) (errorbars show SD, $***$ indicates $p<10^{-10}$). F) Illustration of the heterotypic interaction dynamic, where a an EphB2 (green) cell moves away from an ephrinB1 (red) cell upon interacting with it, exhibiting an increased persistence for about 12 minutes. The illustration also defines the exit angles (see \fref{sfig:stats}A and Appendix).}
    \flabel{fig:stats} 
\end{centering}
\end{figure*}

In what follows, we present our experimental observations confirming the existence of a differential persistence mechanism. We then introduce a minimal theoretical model to numerically study differential persistence, and other cell sorting mechanisms, including differential adhesion and contact inhibition of locomotion (CIL), as well as combinations of them.

\section{Results}

\subsection{Eph-ephrin-mediated differential persistence}

To investigate the motility of individual cells and the physical mechanisms that lead to cell sorting, we performed individual cell tracking (see appendix \ref{cell_tracking_description}). First, we turn to the low-density experiments (\fref{fig:experiments}A). This diluted condition allows the detailed study of individual cell behaviour and pair interactions between same type and different type cells. To check for global differences in persistence, we study the tortuosity $T$ of the cell tracks, which is defined as the relative difference between the total distance travelled $L$ and the straight-line displacement $I$ of a cell $T= \frac{L-I}{L}$, with a ballistic trajectory corresponding to $T=0$ and a trajectory that returns to its starting point corresponding to $T=1$. We observe statistically significant differences between the distributions of tortuosities for EphB2$^+$ and ephrinB1$^+$ trajectories (see \fref{fig:stats}A), $p=9.5e^{-14}$ from a two-sample Kolmogorov-Smirnov (KS) test, indicating that EphB2$^+$ cells tend to move more persistently in this condition.

In order to understand these differences in motility between the two cell types, we looked at pairwise contacts: By using the contact distance of 30 $\mu$m we could identify cells that go from being in contact with a single cell to moving freely without any contacts (groups of three or more cells in contact were excluded from this analysis). By identifying the moment of a contact between two cells in each of the cell tracks, we could analyse the dynamics of the cells after an interaction (see \frefs{fig:stats}B-E). For this, we grouped the post-contact tracks based on the cell (ephrinB1$^+$ or EphB2$^+$) as well as the type of collision: If a cell contacts a cell of the same cell type, the interaction is homotypic (hom), if it interacts with a cell of different type, the interaction is heterotypic (het). 
We now perform a linear fit to the total distance travelled by each group of cells in order to obtain an estimate for their self-propulsion speed $v$ (see \fref{fig:stats}B), which remained constant at $v \approx 1.2$ $\mu$m/min in all conditions, consistent with prior reports on total distance covered \cite{ONeill2016,Kindberg2021}. Fixing the value of $v$, we then fitted the mean-squared displacement of the cells (\fref{fig:stats}C) against the known result for an active Brownian particle in the limit of high Péclet
\begin{align}\label{eq:msd}
MSD(t) = 2 v^2 \tau t + 2 v^2 \tau^2 \left( \exp{-t/\tau} -1 \right),
\end{align}
where $t$ measures the time elapsed since the last contact. Here, we assume that fluctuations in the cell position are negligible when compared to the self-propulsion \cite{bechinger2016active}, allowing us to neglect translational diffusion, which would contribute with an additional linear term in $t$.

The resulting values for the self-propulsion speed $v$ and the persistence time $\tau$ are shown in \frefs{fig:stats}D-E. For the persistence time, we see significant variations, with all the persistence times falling around $\tau$ = 4  min, with the exception of EphB2$^+$ cells, whose persistence time after a heterotypic contact tripled to $\tau$ = 12.7 min. Notably, this increased persistence is independent of self-propulsion speed (see \frefs{fig:stats}D and \fre{fig:stats}E). This analysis reveals that the difference in tortuosity found above (\fref{fig:stats}A) comes entirely from EphB2$^+$ cells, which adopt a transient state of increased persistence after heterotypic contacts with ephrinB1$^+$ cells (see schematic \fref{fig:stats}F). This provides evidence that a differential persistence mechanism is present in these cells and acts during cell sorting. 

In the following, we characterise the cell sorting dynamics in high-density conditions, and we later investigate the effects of persistence through numerical simulations.

\subsection{Heterotypic interactions lead to cell sorting}

Consistent with previous observations \cite{poliakov2008}, sorting was observed in high-density EphB2$^+$-ephrinB1$^+$ cocultures, where EphB2$^+$ clustered together into densely packed islets that were surrounded by a more diluted phase of red-labelled ephrinB1$^+$ cells (see \fref{fig:experiments}C). This phenomenon was absent in the low-density coculture (\fref{fig:experiments}A) and in the high density EphB2$^+$-EphB2$^+$ control (see \fref{fig:experiments}B), where cells remained in a diluted well-mixed state for the duration of the experiment. This indicates that repeated heterotypic EphB2$^+$-ephrinB1$^+$ interactions are necessary for cell sorting, which are favoured in the high cell density conditions.

\begin{figure}[t!]
\begin{centering}
    \includegraphics[width=0.5\textwidth]{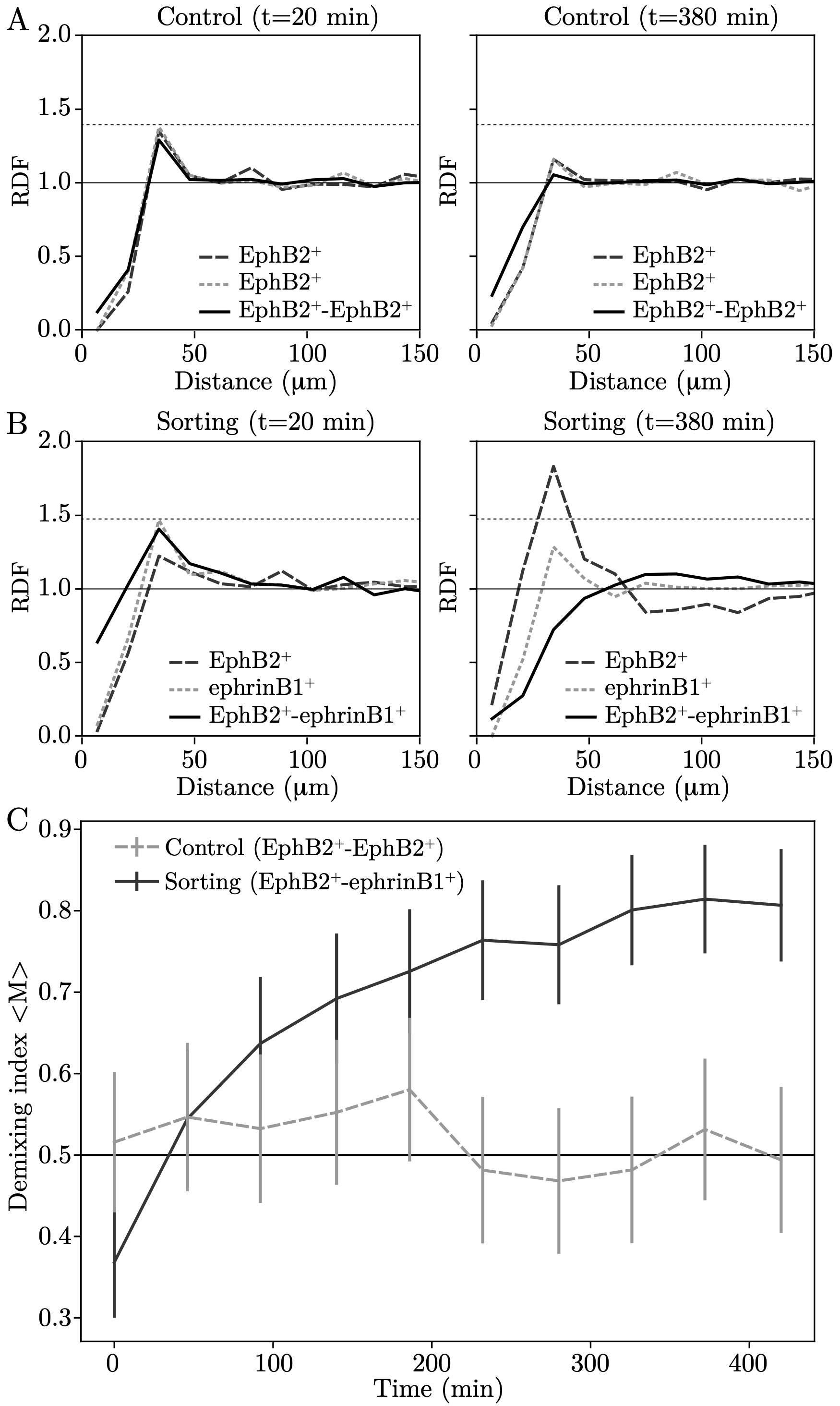} 
    \caption{(A-B) Radial distribution functions (RDFs) for A) the control experiment and B) the sorting experiment. The RDFs labelled as EphB2 and ephrinB1 consider only the distances between cells of the same type and labelling color, while EphB2-EphB2 and EphB2-ephrinB1 RDFs takes into account only cell pairs with different labelling. The reference black line marks the uniform distribution $RDF=1$, while the dashed horizontal reference line marks the height of the initial peaks. C) Demixing index as a function fo time in the high-density cases for the control and segregating experiments. The reference black line marks the perfect mixing value of 0.5.}
    \flabel{fig:rdfs} 
\end{centering}
\end{figure}

From the complete trajectories of individual cells, we computed the radial distribution function (RDF), \ie the probability of finding a cell of a particular type at a certain distance from another cell (see Appendix \ref{RDF_definition} for details). In this way, we constructed the same-type (see~\fref{fig:rdfs}A) and cross-type RDF (see~\fref{fig:rdfs}B). The RDF quantifies spatial dependence of the cell density as a function of the distance from a reference cell (located at $r=0$), hence for a well-mixed, uniform distribution of cells, the RDF has a value close to 1, indicating minor deviations from the average density. Due to volume exclusion, all RDFs (\fref{fig:rdfs}) exhibit a significant depression for $r<2r_\text{cell}\approx 30 \mu$m.

At early times, both control and sorting high-density experiments exhibit a low peak in the RDF at around one cell diameter $\sigma$ = 30 $\mu$m (indicated by dashed lines in \frefs{fig:rdfs}A-B), as expected from the near-confluent initial condition. In the control case, the peak is slightly reduced over time, with no change in the short distance depression. However, for the EphB2$^+$-ephrinB1$^+$ coculture we see a strengthening in the EphB2$^+$-EphB2$^+$ RDF peak with time, while the cross-type RDF loses this peak completely, as a result of the effective repulsion between distinct types during cell sorting. The EphB2$^+$ peak is much more pronounced than the ephrinB1$^+$ peak, indicating stronger clustering among the EphB2 cells. Moreover, at $t=400$ min (see \fref{fig:experiments}) the total area covered by ephrinB1$^+$ (red) cells was about 50\% higher than that covered by EphB2$^+$ (green) cells. Considering that the experiments are all plated with equal amounts of EphB2$^+$ and ephrinB1$^+$ cells, these results are consistent with the observation that islands of EphB2$^+$ (green) cells are formed while being surrounded by a dilute phase of ephrinB1$^+$ (red) cells (see \frefs{fig:experiments}B-C).   

As a summary statistic to quantify the cell sorting dynamics, we measured the demixing index. For the $i$th cell of a given type, the demixing index $M_i$ measures the fraction of its nearest neighbour cells that are of its same cell type, i.e.,
\begin{equation}
M_i = \frac{\text{Number of same-type neighbours}}{\text{Total number of neighbours}} .
\end{equation}
Here we define the neighbourhood of a cell as all surrounding cells that lie closer than 30 $\mu$m from it, measured centre to centre. We then averaged the $M_i$ over all cells of a given type to obtain an average demixing index $\ave{M}$. A perfectly mixed experiment corresponds to $\ave{M} = 0.5$, whereas a completely demixed (sorted) state approaches $\ave{M}=1$. This limit would be reached if each cell type formed isolated clusters, with no contact between different-type clusters.

Quantification of the average demixing index in the sorting experiment revealed an initial ``overmixing'' with a value $\ave{M} < 0.5$ (see solid curve in \fref{fig:rdfs}C). This is due to correlations in the initial cell distribution, possibly caused by not all cell bonds being broken during resuspension and plating. Over time, the average demixing index in the sorting condition rises rapidly to reach a value of $\ave{M}\approx0.8$ after 200 minutes, indicating a high degree of cell sorting. In contrast, the control setting fluctuates around a value of $\ave{M}=0.5$, for the duration of the experiment (see dashed curve in \fref{fig:rdfs}C). 

In the following, we address the existence of other cell sorting mechanisms in the experiments.

\subsection{Contact inhibition of locomotion and adhesion}

In addition to changes in persistence, EphB2$^+$ (green) cells appear to change their direction of motion upon contact with ephrinB1$^+$ (red) cells. This behaviour lies within the general mechanisms of CIL \cite{stramerMechanismsVivoFunctions2017}, which amounts to a directed reorientation of the leading edge of the cell after interacting with another cell. To analyse this directed motion, we measured the exit angles, defined as the angle between the displacement vector of a cell after contact and the line connecting the centres of both interacting cells at the time of contact (see schematic \fref{fig:stats}F and Appendix \ref{exit_angle_definition}). Thus, an exit angle of 0$^\circ$ describes a contact where the cell moves directly away from its contact partner.

The distribution of exit angles (see \fref{sfig:stats}A) in the high-density EphB2$^+$-ephrinB1$^+$ case shows a narrow distribution, with the majority of angles concentrated in the range $[0,60]^\circ$, compared to the broader distribution for the control.
Seen as symmetric distributions around 0$^\circ$ they differ in their variance, which is 117$^o$ for the homotypic case, and 93$^\circ$ for the heterotypic case ($p=0.0003$ from a KS-test). This indicates a more controlled directional motion by EphB2$^+$ upon heterotypic contact with ephrinB1$^+$ cells, compared to other cell type interactions, suggesting that mechanisms for heterotypic CIL are acting between EphB2$^+$ and ephrinB1$^+$ cell types, which may also be contributing to cell sorting.

In addition to CIL, cells may exhibit differential adhesion, which, as mentioned above, has been proposed as a mechanism for cell sorting. To investigate whether differences in adhesion exist in conditions of pairwise interactions, we investigated whether the duration of cell-cell contacts varied depending on the type of interactions (\fref{sfig:stats}B), and observed an increase of $10-25\%$ for heterotypic contacts. Despite the longer duration of heterotypic contacts, they remained finite, lasting on the order of 20 minutes. This suggests that even if present, differential adhesion is not strong enough to form durable bonds between cells nor to drive cell sorting in our experiments. 

In summary, we have found strong experimental evidence of cell sorting in high-density EphB2$^+$-ephrinB1$^+$ cocultures, which appears to be driven by differences in differential motile behaviour upon heterotypic contacts. Notably, this change in motility of EphB2$^+$ cells does not originate from the change in movement speed observed in other systems \cite{beatrici2011cell,mehes2012collective}, but only changes in persistence. Additionally, measurements of the exit angle revealed the presence of heterotypic CIL, whereby EphB2$^+$ cells tend to orient their motion away from ephrinB1$^+$.  In order to connect these cell-level mechanisms with the global sorting behaviour, and understand the effect of persistence in the sorting dynamic, we turn to numerical simulations, which allow us to isolate these effects and study them over a broad range of parameters.

\begin{figure*}
\begin{centering}
    \includegraphics[width=1\textwidth]{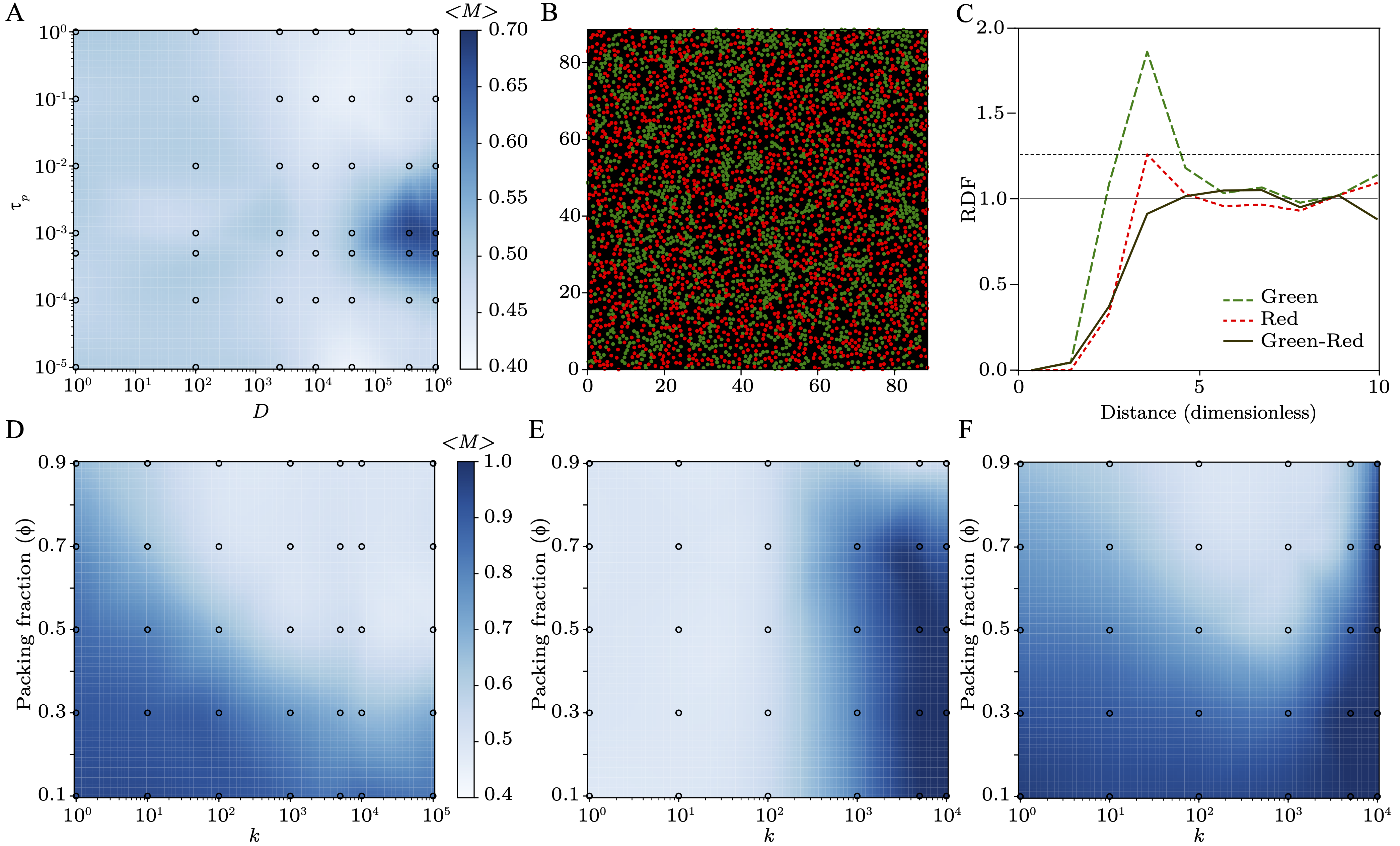} 
    \caption{Cell sorting for (A-D) different persistence, (E) differential adhesion, and (F) a combination of both mechanisms. (A-C) Differential persistence results, showing (A) Demixing index for varying $\tau_p$ and $D$, B) final state of a simulation in the demixing (blue) region in (A) with $\phi=0.4$, $D=3.6\times10^5$, $\tau_p=0.002$, $\textrm{Pe}=10^3$, $k=10^3$, and C) its corresponding RDF, where the reference black line marks the uniform distribution $RDF=1$ and the dashed horizontal reference line marks the height of the peak for the red-red RDF. (D-F) Demixing index on the packing fraction $\phi$ and repulsive potential strength $k$ parameter space, for the simulations of D) differential persistence, E) differential adhesion, and F) for combined differential adhesion and differential persistence. Colorbars of (E) and (F) are the same as in (D). The values of the demixing indices are always obtained from the final states of the simulations} 
    \flabel{fig:mixings}
\end{centering}
\end{figure*}

\subsection{Model and simulations}

To study the different mechanisms for cell sorting, we propose a simplified model of interacting soft particles that behave as active Brownian particles (ABPs) \cite{romanczuk2012active}, with constant self-propulsion speed $v$. In two dimensions, the equations of motion for the position $\ve{r}_i$ and orientation $\phi_i$ of the $i$th particle are given by
\begin{align}
\dot{\ve{r}}_i(t) &= v \hat{\ve{n}}_i - \sum_j \boldsymbol{\nabla}_i U(r_{ij}) + \sqrt{2 D_T} \boldsymbol{\eta}_{T,i}(t) \\
  \dot{\varphi}_i(t) &= F_\text{CIL} +  \sqrt{2 D_{R,\alpha}} \,\eta_{R,i}(t) \, ,
\end{align}
where $\hat{\ve{n}}_i = (\cos{\varphi_i},\sin{\varphi_i})^\top$, and $U(r_{ij})$ corresponds to the interaction potential between cells, $r_{ij}=|\ve{r}_i-\ve{r}_j|$ is the pairwise distance between cells, $D_T$ is the translational diffusion constant, and $D_{R,\alpha}$ is the cell-type-dependent rotational diffusion constant, with $\alpha \in \{\text{green}, \text{red}\}$. The inter-particle CIL effect that enforces a directional change of the $i$th cell is included as $F_\text{CIL}$, which is implemented numerically as an instantaneous anti-alignment between the two interacting cells. The elements of the vector $\boldsymbol{\eta}_{T,i}$ and $\eta_{R,i}$ are taken from independent zero mean and delta-correlated normal distributions. To reduce the number of free parameters we nondimensionalise the equations of motion by using the cell diameter $\Delta x= \sigma$ as the length scale and defining our timescale $\Delta t = \frac{\sigma^2}{D_T}$ via the translational diffusion constant. Despite the fact that translational diffusion plays a minor role in our experiments, this choice facilitates the connection to the wider physics literature on active matter. The biological conditions can then be recovered in the large Péclet number (Pe) limit. With this, we arrive at the nondimensional equations
\begin{align}
\dot{\ve{r}}_i(t) &= \text{Pe\,} \hat{\ve{n}}_i - \sum_j \boldsymbol{\nabla}_i \tilde{U}(r_{ij}) + \sqrt{2} \boldsymbol{\eta}_{T,i}(t) \label{eq:model1} \\
  \dot{\varphi}_i(t) &= \tilde{F}_\text{CIL} +  \sqrt{2 D_\alpha} \,\eta_{R,i}(t) \label{eq:model2} \, , 
\end{align}
where the self-propulsion speed is given by the Péclet number $\text{Pe} =\frac{v\sigma}{D_T}$ and the effective rotational diffusion constant was defined as $D_\alpha = \frac{D_{R,\alpha} \sigma^2}{D_T} $. For the interaction potential, we consider a harmonic potential, which is relatively soft and thus captures the capacity of cells to deform and squeeze past each other (a Lennard-Jones potential was also considered and gives qualitatively the same results).
The resulting force term for one cell is then 
\begin{align}\elabel{eq:force}
- \sum_j \boldsymbol{\nabla}_i \tilde{U}(r_{ij}) = k \sum_j \frac{\ve{r}_i-\ve{r}_j}{r_{ij}} (r_{ij}-1) \, \Theta (\sigma_\text{max} - r_{ij}) \,,
\end{align}
which has two free parameters: the strength of the potential $k$ and its range $\sigma_\text{max}$. Purely repulsive interactions are obtained by setting $\sigma_\text{max}=1$, while attractive interactions can be accounted for by setting $\sigma_\text{max} > 1$. In particular, we will use $\sigma_\text{max}=1.1$ when we want to model short-range adhesive forces between cells.

Within this model we will consider two species (green and red), which may respond differently to an interaction with a cell of the same or different type, mimicking the EphB2$^+$-ephrinB1$^+$ behaviour. Differential persistence is implemented as a temporary change of the rescaled diffusion constant $D_\alpha$: When a green and a red cell come into contact (i.e. $r_{ij}<1$), the cells enter activated states with diffusion constants $D_\text{green}^+, D_\text{red}^+$, which last for a finite persistence time $\tau_p$. In accordance with our experimental findings in \fref{fig:stats} we assume identical baseline constants $D \equiv D_\text{green} = D_\text{red}$ and lower rotational diffusion (higher persistence) for green cells, $D^+ \equiv D_\text{green}^+< D_\text{green}$, with no change in red cells, $D_\text{red}^+ = D_\text{red} = D$. 

The CIL mechanism is implemented by changing the direction of self-propulsion of both interacting cells, resulting in both cells moving away from the common point of contact. For details of the numerical implementation see Appendix~\ref{app:numerical_sims}.

The set of coupled stochastic differential equations (\ref{eq:model1}) and (\ref{eq:model2}) is implemented numerically using an Euler–Maruyama integration scheme. In the following, the parameters used in all simulations are given in length units of the cell diameter $\sigma$ and time units of $\frac{\sigma^2}{D_T}$.

\subsection{Differential persistence induces cell sorting}
To understand the effects of differential persistence in cell sorting, we first study this mechanism in isolation by simulating two species of cells, with the green species entering into a high-persistence state upon contact with red cells. 
First, we consider a relatively low packing fraction of $\phi = 0.4$, $\phi$ being the fraction of the simulated area covered by cells, and restrict ourselves to purely repulsive cells ($\sigma_\text{max}=1$ in Eq.~\eref{eq:force}). To ensure that we are in the high-Péclet regime, we set $\textrm{Pe}=k=10^3$ and look for the optimal configuration to induce sorting by varying $D$, $D^+$ and $\tau_p$. 

We note that $1/D^+$ corresponds to a persistence time, i.e., the time it takes for a cell to randomize its orientation when in the high persistence state, while $\tau_p$ corresponds to the time interval for which the cell remains in the high persistence state. Here we focus on the limit where $1/D^+ \gg \tau_p$ by setting $D^+=0$, i.e., cells move ballistically while in the activated state. The parameter to explore then reduce to $\tau_p$ and $D$.

The corresponding $D - \tau_p$ parameter space  (\fref{fig:mixings}A) shows that a transient differential persistence can produce sorting on its own. The values shown are recorded at $t = 5 \, \Delta t$, the time at which all demixing indices have reached steady-state values. Here, nearest neighbours are defined as all cells that are within $1.1\sigma$ of the reference cell. Demixing grows larger with $D$, with substantial demixing ($\ave{M}>0.6$) occurring for $D>4 \cdot 10^4$. The resulting demixed configuration is shown in \fref{fig:mixings}B, where a clear asymmetry between green and red cells is observed, consistent with experimental observations (see \fref{fig:experiments}C). Here, green cells undergo repeated persistence changes forming high-density clusters surrounded by more diluted red cells. The radial distribution function for the demixed state (\fref{fig:mixings}C) shows peaks for both types of cells at a distance of one cell diameter; however, with a much more pronounced peak for green cells and a suppressed green-red correlation throughout, as observed experimentally (see \fref{fig:rdfs}C). 

To provide some comparison to experimental values, we may take into account that at $\textrm{Pe}=10^3$ the simulated cells propel themselves $10^3$ cell diameters ($\sigma=30$ $\mu \text{m}$) for each simulation time unit. From comparing this to the experimentally observed speed of $v = 1.2$ $\frac{\mu \text{m}}{\text{min}}$, or $v = 0.04\, \sigma /\text{min}$, we obtain that a single simulation time unit corresponds to $\Delta t = 2.5\times 10^4 \, \text{min}$ (see Appendix \ref{matching_parameters}). In physical units, a simulation diffusion constant of $D = 3.6 \times 10^5$, which produces demixing of $\ave{M}=0.68$, amounts to a rotational diffusion constant of $D_R = 14.4 \,\frac{1}{\text{min}}$, which is substantially larger than the experimental estimate of $D_{R,\text{exp}} = \frac{1}{\tau} = 0.25 \frac{1}{\text{min}}$. This discrepancy can be attributed to our theoretical approximation of a cell via a short range potential, when in reality, cells are capable of longer range sensing through filopodia, small filament-like protrusions essential for cell-cell communication \cite{ruhoff2024filopodia}, which are no visible under bright-field imaging. Filopodia allow cells to explore their surrounding without a effective change in direction of motion, while numerical cells can only explore their environment and contact cells through changes in directions, hence the increased rotational diffusion required for cell sorting.

The phase diagram (\fref{fig:mixings}A), revealed that there is an optimal value for the duration of the persistent state $\tau_p^{\text{opt}}=\tau_p = 10^{-3}$ for sorting to occur. From a value of $\textrm{Pe}=10^3$ and the optimal duration $\tau_p^{\text{opt}}=\tau_p = 10^{-3}$, the path length of the high-persistence state is of the order of one cell diameter. For larger values of $\tau_p$, cells update their persistence less frequently and, thus, are able to move out of a cluster of same-type cells, reducing the demixing. Conversely, for smaller values of $\tau_p$, the persistence length becomes small compared to the size of the cell and thus the high persistent motion is comparable to its normal dynamics. 

The experimental results (\frefs{fig:stats}D-E) for activated, high-persistence cells show a speed of $v$ = 1.2 $\frac{\mu \text{m}}{\text{min}}$ and a persistence duration of $\tau$ = 12.7 $\text{min}$, resulting in a mean path length of $v \cdot \tau = 15.2$ $\mu \text{m} = 0.5 \sigma$, or half a cell diameter. By comparison, the mean path length in the optimal region is $\textrm{Pe} \cdot \tau_{opt} = 1 \sigma$, with simulations with half this path length showing similar values of demixing. This shows that the experimental conditions lie near the predicted parameters for optimal demixing.
 
Further numerical analyses varying $\textrm{Pe}$ showed that the optimal value of $\tau_p$ changes inversely with $\textrm{Pe}$, indicating that the effect does depend only on the mean path length. To stay close to the optimal value, in the following we therefore fix $\tau_p = 2/\textrm{Pe}$ unless stated otherwise.

Now, we turn our attention to the effect that packing fraction (i.e. cell density) has on the sorting. There is an interplay here between the nominal packing fraction, which assumes that each cell covers an area of $\frac{\pi}{4}$, and the strength of the repulsive force, since the latter governs cell overlaps, which reduce the effective packing fraction. The effects of varying these parameters on the demixing index are shown in \fref{fig:mixings}D, where we used the optimal parameter values for cell sorting $D=3.6\cdot 10^5$ and $\tau_p = 0.002$ found above. The demixing index increases for lower packing fractions and lower potential strength, and is almost non-existent as we approach full coverage with hardcore cells. This indicates that the capacity to move freely and squeeze past cells is essential for the differential persistence mechanism to work at high cell densities.

In summary, our simulations qualitatively replicate the demixed state observed in high-density  EphB2$^+$-ephrinB1$^+$ experiments (\frefs{fig:experiments} and \fref{fig:rdfs}). Through differential persistence, cells segregate into dense clusters of green cells surrounded by a less dense ``gas'' of red cells, exhibiting a RDF consistent with experiments. While there is a discrepancy in the mean path length between simulation and experiment, the simulation match the experimental high-persistence path length ($v \cdot \tau^+ = 0.52 \sigma$), and show that its value is optimal for sorting. 
However, the observed demixing of $\ave{M}=0.65$ remains below the value of $\ave{M}=0.8$ observed in the high-density EphB2$^+$-ephrinB1$^+$ experiments, suggesting that persistence change may not be the only mechanism acting during cell sorting. In the following, we investigate numerically how differential adhesion and CIL alone, and in combination with differential persistence influence cell sorting.

\subsection{Effect of differential adhesion}

While our experiment did not show significant evidence of differential adhesion, cells in general can form durable membrane bonds that strengthen over time \cite{garcia2015physics}. Acknowledging this mechanism's potential dominance in other contexts, we examined the combined impact of differential persistence and adhesion on sorting dynamics. In our model, adhesion is represented by extending the harmonic potential's range to an adhesive region ($\sigma_\text{max}=1.1$), where the potential's strength dictates cell rigidity and adhesive force. We modelled differential adhesion by limiting adhesive interactions to same-type cell contacts.

Differential adhesion alone may lead to cell sorting, evidenced by increased demixing (see $k$-$\phi$ parameter plane for $\textrm{Pe}=1000$ and $D=200$, \fref{fig:mixings}E), which scales with the adhesive force strength $k$, peaking at $k \geq \textrm{Pe} =1000$. This sorting effect occurs across various packing fractions, except around $\phi=0.9$, where high $k$ values approach a jamming transition \cite{henkes2011active}, causing the system to freeze in its initial state. Although most sorting occurs within the first 0.25 time units, there is a gradual, long-term ripening, which would be too slow to observe experimentally, where other mechanisms and temporal changes are likely to take place.
Additional simulations reveal that a higher rotational diffusion constant enhances cell sorting. This enhancement occurs because increased diffusion leads to more randomized cell motion, making it more difficult for cells to escape the attractive potential and thereby intensifying the effect of differential adhesion.

Combining differential adhesion with differential persistence has significant effects in the sorting process. If green cells are allowed to increase their persistence by reducing $D$ from $D=3.6 \cdot 10^5$ to $D=10^2$ upon heterotypic contact, significant demixing is observed at lower $k$ values (\fref{fig:mixings}F) compared to the purely adhesive scenario (\fref{fig:mixings}E). This results from the fact that a weak potential (low $k$) limits adhesion effectiveness, while the persistence effect remains relevant.

\begin{figure}
\begin{centering}
    \includegraphics[width=0.5\textwidth]{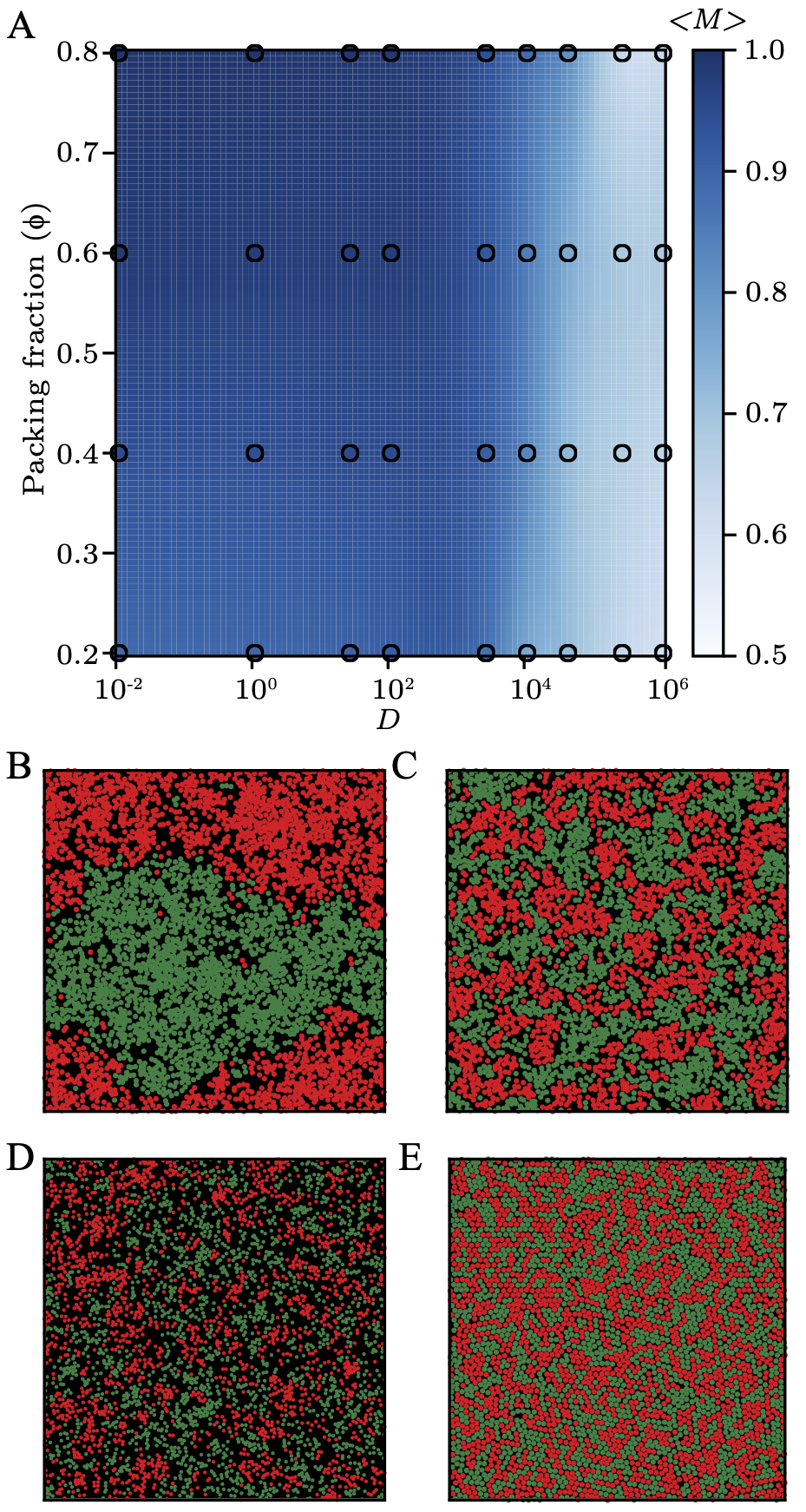} 
    \caption{Numerical results for CIL-based sorting at $\textrm{Pe}=k=10^3$ for different values of rotational diffusion constant $D$ and packing fraction $\phi$. A) The demixing index in the $D$-$\phi$ plane. Final configurations at B) $\phi=0.8$, $D=1$, C) $\phi=0.8$, $D=2500$, D) $\phi=0.4$, $D=1$, and E) $\phi=0.8$, $D=10^6$.} 
    \flabel{fig:mixingExampleCIL}
\end{centering}
\end{figure}

\subsection{Heterotypic contact inhibition of locomotion}

To isolate the effects of CIL in cell sorting, we consider the model with no adhesion nor persistence changes. Fixing $\textrm{Pe}=k=10^3$, we consider non-differential heterotypic CIL, which is implemented as an instantaneous change in the direction of movement of both interacting cells upon heterotypic (green-red) interactions. The change in direction is symmetrical (non-differential), so that both cells anti-align, moving in opposite directions and away from the point of contact. Unlike the commonly studied alignment interaction, anti-alignment interactions break the symmetry of the system under shared rotations of velocity and space, which has been shown to result in cell clustering and ordering \cite{bertrand2024clustering}. Our implementation of CIL is an instantaneous and deterministic limit of the real biological dynamic, where CIL requires a finite amount of time of contact for cell polarisation and motion to take place, and where there are finite probabilities for the effect to trigger in both heterotypic and homotypic contacts \cite{taylorCellSegregationBorder2017}. 

The phase-diagram (\fref{fig:mixingExampleCIL}A) shows that CIL induces cell sorting for the entire parameter range, with an increased demixing in dense conditions (high packing fraction), under high frequency of collisions. Unlike differential persistence, CIL requires relatively low rotational diffusion constants ($D<10^4$) for optimal performance, as high diffusion negates directional changes. At $\phi=0.8$ and $D=1$, CIL not only achieves local sorting but also global phase separation with distinct red and green phases (\fref{fig:mixingExampleCIL}B). We note that these clusters do not form perfect crystalline (hexagonal) configurations as typically observed in standard MIPS (see \fref{sfig:mips}) \cite{hopkins2023motility}. Cluster size, influenced by the cell path length ($\textrm{Pe}/D$), varies with rotational diffusion; for $D=50$, finite-sized clusters persist for the duration of the simulations, indicating micro-phase separation (\fref{fig:mixingExampleCIL}C) \cite{van2019interrupted}. Lower packing fractions ($\phi=0.4$) disrupt global clusters at high persistence $D=1$ (see \fref{fig:mixingExampleCIL}D), reminiscent of the critical density conditions for motility-induced phase separation in repulsive active particles \cite{CatesTailleur2015}. However, at very high diffusion constants ($D=10^5$), the sorting mechanism is hindered by the effectively diffusive behaviour of cells, leading to higher coverage and reduced cell overlap (\fref{fig:mixingExampleCIL}E).

These results were obtained with purely symmetrical interactions, and thus there were no symmetry-breaking between the two cell types. Introducing differential persistence--where only green cells increase their persistence upon heterotypic contact--leads to distinct cell distributions (see \fref{fig:clusteringCILdp}). 
By varying the relation between $D$ and $D^+$, we could control the difference in average cluster size $\ave{N}$ between green and red cells (see \fref{fig:clusteringCILdp}A). For a fixed value of $D^+$, a higher differential persistence ($D^+ < D$), achieved by increasing $D$, resulted in a breakup of the single large green cluster into well-defined smaller groups of green cells surrounded by a ``gas'' of red cells (see bottom-left corner in \fref{fig:clusteringCILdp}A and \frefs{fig:clusteringCILdp}B-C).
The configuration is inverted by fixing $D=100$, and lowering the differential persistence ($D^+ > D$). This results in the formation of red clusters surrounded by green cells (see differences between \frefs{fig:clusteringCILdp}D-C). 

\begin{figure}
\begin{centering}
    \includegraphics[width=0.5\textwidth]{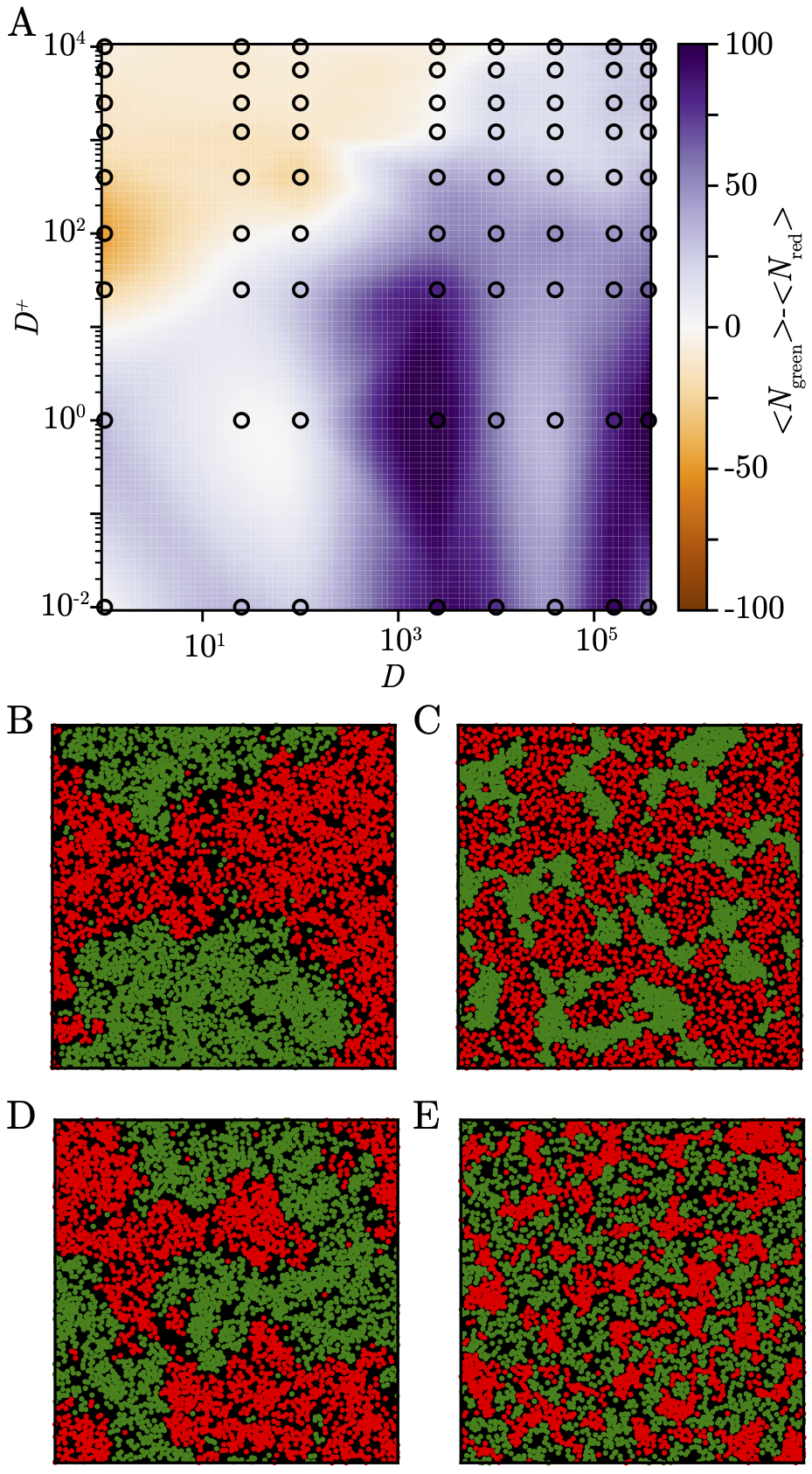} 
    \caption{Numerical results for CIL-based sorting with differential persistence at $\textrm{Pe}=k=10^3$, $\tau=0.002$ for different values of $D$ and $D^+$ at a packing fraction of 0.8. A) Shows the difference between the average cluster sizes of the green cells $\ave{N_{green}}$ and the red cells $\ave{N_{red}}$. Final configurations at B) $D=100$, $D^+=0.01$, C) $D=10^4$, $D^+=0.01$, D) $D=100$, $D^+=100$, and E) $D=100$, $D^+=2500$.} 
    \flabel{fig:clusteringCILdp}
\end{centering}
\end{figure}

This combined mechanism enables both global and micro phase separation, where the breaking of symmetry is controlled by differential persistence. To test whether CIL alone could reproduce the experimental observations, we simulated a variation of the CIL model, where only green cells change their direction upon heterotypic contact, but this setup failed to induce any sorting, let alone the clustering structure seen in our experiments.
Interestingly, by considering differential persistence in the dense phase ($\phi=0.9$, see \fref{sfig:singlefile}A) and strong repulsive potential, the system may reach a phase with demixing index $\ave{M}<0.5$. This is achieved by the formation of a single-file phase, where cells form a hexagonal lattice of alternating red and green cells, resulting in a demixing index lower than 0.5 (see \fref{sfig:singlefile}B and discussion in Appendix~\ref{app:single_file}).

In summary, our findings indicate that CIL effectively facilitates strong cell sorting and can lead to global phase separation. Cell motility plays a crucial role in determining the cluster's final structure, where higher persistence correlates with larger clusters and system-wide bands. Unlike the isolated differential persistence effect, CIL proves more efficient at higher densities and lower rotational diffusion constants. However, only when combining CIL with differential persistence could we reproduce the type of cell sorting observed in Eph$B2^+$-ephrinB1$^-$ cocultures (\fref{fig:experiments}A), which strongly suggests that differential persistence is an essential mechanism in this system.

\subsection{Temporal dynamics of cell sorting}

All mechanisms studied here--differential persistence, differential adhesion and CIL, and combinations of them--are able to produce long-term cell sorting. However, the details of the final configuration depend on the details of the interactions \cite{beatrici2017mean}. Moreover, for biological systems, the timescales and speed of sorting are relevant \cite{franke2022cell}, as these processes usually take place in narrow time windows during embryonic development. 
To study the sorting speed, we focus on the onset of the sorting for different combinations of persistence, adhesion and CIL effects. The baseline for this is a system of purely repulsive cells at $\textrm{Pe}=k=10^3$, without differential persistence or CIL effects. The rotational diffusion constant is set to $D=3.6 \cdot 10^5$, as we have seen sorting for all three mechanisms at these parameter values. 

When comparing the sorting dynamics for various mechanisms, we observe that simulations typically reach a steady state after about $5 \, \Delta t$. 
As we want to study the sorting dynamics in the timescale of the experiments, we focus our attention in the early stage dynamics, up to $0.05 \, \Delta t$ (equivalent to about 1000 min in the experiments), where the vast majority of the demixing takes place. 
At low densities ($\phi=0.2$), both differential adhesion and persistence achieve similar sorting efficiencies, nearing $\ave{M}=0.8$  (see \fref{fig:mixingComparison}A). However, differential CIL reaches a much lower value of $\ave{M}=0.6$.
The combination of differential persistence and CIL leads to significantly higher sorting ($\ave{M}=0.88$) in a short time frame compared to the other mechanisms. 

The dynamic is altered significantly at high densities ($\phi=0.6$), where differential persistence alone leads to less demixing, reaching $\ave{M}=0.6$ (see \fref{fig:mixingComparison}B). Both differential CIL and adhesion show slightly reduced demixing compared to the low-density case (\fref{fig:mixingComparison}A), due to the longer time needed for densely packed cells to demix. However, the combination of differential persistence and adhesion, which excelled at low densities, performs less effectively at high densities, reaching only $\ave{M}=0.67$ within the time frame observed. In contrast, the combination of differential CIL and adhesion achieves the highest demixing ($\ave{M}=0.78$),
matching our high-density experimental observations in \fref{fig:rdfs}C, and showing that differential persistence significantly accelerates the cell sorting dynamics.

\begin{figure}
\begin{centering}
    \includegraphics[width=0.5\textwidth]{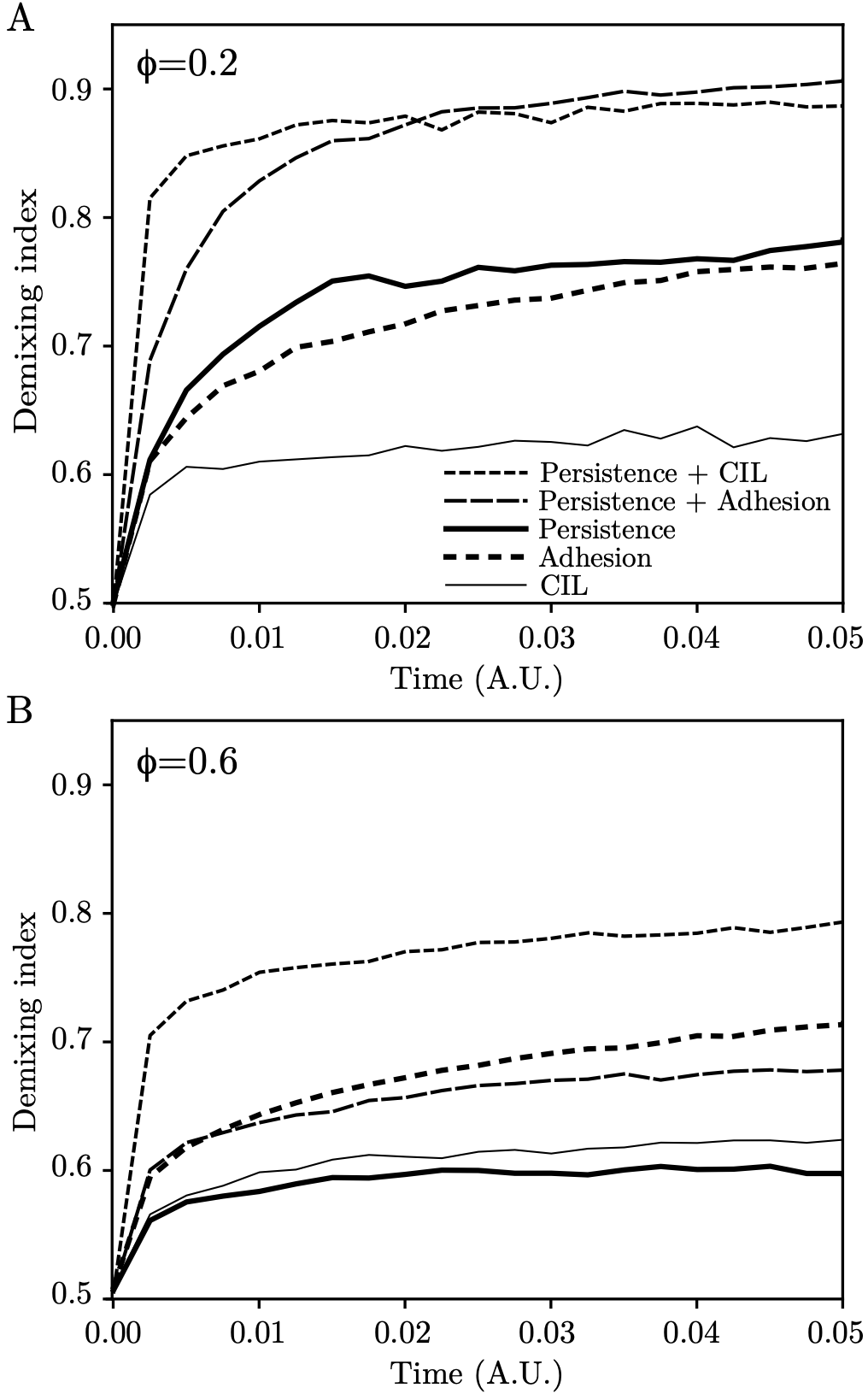} 
    \caption{Demixing index over time for several mechanisms for A) low packing fraction $\phi=0.2$ and B) high packing fraction $\phi=0.6$, averaged over ten realisations.} 
    \flabel{fig:mixingComparison}
\end{centering}
\end{figure}

\section{Conclusion and discussion}

In this work, we investigated how differences in cell-cell interactions may lead to cell sorting in binary mixtures. Analysing cocultures of EphB2$^+$ and ephrinB1$^+$ HEK293 cells, we found strong evidence for the existence of differential persistence, a previously unexplored mechanism for cell sorting, where one cell type (EphB2$^+$) enhances its self-propulsion persistence upon contacting cells of a different type (ephrinB1$^+$). In addition to differential persistence, we found signs of heterotypic CIL, evidenced by the exit angles of cells after heterotypic interaction, consistent with previous studies using the same cell line \cite{poliakov2008regulation}. However, we found no signs of differential adhesion, as homotypic contact durations were not significantly longer than heterotypic ones. Notably, the observed change in persistence was found to be decoupled from cell speed, in contrast to previous observations in other cellular systems \cite{maiuri2015actin}. This is consistent with studies showing no increase in total migratory distance in sorting versus non-sorting cell mixtures \cite{ONeill2016,Kindberg2021}. Moreover, this suggests the existence of molecular mechanisms by which cells may regulate persistence independently from cell speed.

Using active Brownian particle simulations, we investigated the contributions of differential persistence, differential adhesion, and differential CIL to cell sorting. We found that each mechanism can drive sorting within specific parameter ranges: differential persistence alone works well at low densities due to longer free paths, while high densities favour differential CIL because of increased collisions. The combination of differential persistence and CIL drastically reduced the characteristic time and enhanced the degree of cell sorting, with the differential persistence controlling the asymmetry and degree of fragmentation of the final clusters. When comparing the sorting onset across mechanisms, this combined mechanism proved most effective at high densities, aligning qualitatively with our experimental observations, though not precisely in parameter values. This difference likely stems from our model's simplification of complex factors like cell shape and sensing mechanisms that include long-range sensing through filopodia. Moreover, our numerical results show that there is an optimal duration of the persistent state that enhances sorting, and that the cellular behaviour in the experiment is compatible with this optimal state.

In summary, our study highlights differential persistence as a crucial mechanism capable of independently driving cell demixing under appropriate conditions, and likely used alongside other mechanisms to expedite sorting. This mechanism may apply to a wider variety of systems and cocultures \cite{mehes2012collective,song2016dynamic}, and may explain the sorting behaviour observed in other theoretical models, where particles have correlated directions of propagation \cite{belmonte2008self}. It is worth noting that the differential persistence mechanism contrasts with previous findings where active particles caused clustering of passive ones through active avoidance between particles of different types \cite{stenhammar2015activity}, indicating a distinct underlying mechanism. Alternative theoretical approaches have been used to investigate the distinct mechanisms involved in cell sorting, such as phase-field models \cite{graham2024cell}, where a binary mixture of extensile and contractile particles also leads to sorting, with maximum demixing values around 0.7. They, however, take the view that CIL acts solely as a hindrance to movement due to local particle density, thus behaving more like a MIPS mechanism \cite{CatesTailleur2015}. Thus, future challenges include the development of a comprehensive multi-species theoretical framework that accounts for persistence differences and anti-alignment interactions in active systems.

\appendix

\section{Culture preparation}
\label{coculture}

For cell sorting experiments, the cells were first dissociated in Accutase and passed through 0.45 $\mu$m filters to ensure that any preexisting bonds between the cells were broken up before the beginning of observation. After plating, cells were left to settle for around 1 hour before imaging. Time-lapse images were taken every two minutes for over 6.5 hours using a DeltaVision microscope with 10$\times$ magnification, resulting in a spatial resolution of 3.34 $\mu$m$/$pixel. 

\section{Cell tracking}
\label{cell_tracking_description}

For the cell tracking, we used TrackMate 7 \cite{ershov2022trackmate} with the Laplacian-of-a-Gaussian spot detection method, with an estimated spot size of 8 pixels. To track the spots, we used the linear assignment method, with a maximum displacement of 10 pixels per time step. We allowed for spots to disappear for up to 5 frames to mitigate weaknesses in the tracking. The missing spots were then inserted into the tracks through linear interpolation. 

\subsection{Radial Distribution Function}
\label{RDF_definition}

The radial distribution function was calculated in the standard way \cite{bordeu2020effective}, by sweeping over all the cells found at a given time, and measuring the distances between them and every other cell of the same type and of a different type, separately. To avoid distortions due to the rectangular field of observation, we normalised the histograms of pairwise distances by the histogram obtained for a uniform distribution of particles.

\subsection{Exit Angles}
\label{exit_angle_definition}

As we do not have access to information on cell polarity from the experiments, we can only infer the instantaneous orientation of the cells from their displacements after a contact. For the measurement of the exit angles, we defined contacts as two cell centres being within 9 pixels or 30 $\mu$m from each other. We track the number of contacts a cell has at each point in the experiment. When such a contact breaks, we track the cell over the next 10 min, and define its instantaneous direction based on its displacement after that time (see \fref{sfig:stats}A). The exit angle is then defined as the angle between its instantaneous direction and the vector connecting the centres of both cells at the end of the contact. To avoid problems with cells leaving the field of measurement or getting lost in tracking we restrict the analysis to cells that are tracked for the full duration of the experiment, and we exclude contacts that only last a single time step for the same reason. Finally, we restrict the analysis to cells that are only in contact with one cell before separation, to avoid other cells distorting the trajectory. The same selection criteria are applied when studying the motility of cells after contact. 

\section{Numerical implementation}\label{app:numerical_sims}

Solution of the stochastic differential equations was implemented using the Euler-Maruyama scheme. To speed up the calculation of the interactions, a segmentation of the simulation space and updating neighbourhood lists were used \cite{rapaport2004art}. Random numbers are generated using the xorhsiro256 random number generator.

The initial positions of the cells are chosen randomly, in combination with a gradient descent optimiser that minimises overlaps between cells. Simulations were run with a fixed time step of $10^{-5}$. Usually, the number of cells was chosen to be 2000 of each species, which is roughly double the number we see in the high-density experiments.

\section{Matching experiment and simulations}
\label{matching_parameters}
In order to match the dimensionless simulation parameters to the measured experimental values, we need two conditions, which fix both the length- and the time-scale of the simulations. For this, we choose the cell diameter and the self-propulsion speed because they are both easily accessible from the experimental data. The two conditions are therefore 
\begin{align}
    \sigma \demand \sigma_{exp} = 30 \mu \text{m} \, ,
\end{align}
and 
\begin{align}
    v \demand v_{exp} = 1.2 \frac{\mu \text{m}}{\text{min}} \, .
\end{align}
Other observables then follow from the definition of the dimensionless parameters: Recall that we had defined the unit simulation time as $\Delta t = \frac{\sigma^2}{D_T}$, and the Péclet number as $\textrm{Pe} = \frac{v \sigma}{D_T} = \frac{v}{\sigma} \Delta t$. Rewriting the last condition, we find
\begin{align}
    \Delta t &= \textrm{Pe} \frac{\sigma}{v}  \\
    &=  \textrm{Pe} \frac{\sigma_{exp}}{v_{exp}} \, ,
\end{align}
where, in the last line we inserted the matching conditions. We have $\frac{v_{exp}}{\sigma_{exp}} = 0.04 \frac{1}{\text{min}}$, so for a simulation with $\textrm{Pe} = 1000$ a single time unit corresponds to $\Delta t = 2.5 \cdot 10^4$~min.

Furthermore, we can match the dimensionless rotational diffusion constant $D = \frac{D_{R} \sigma^2}{D_T} = D_R \Delta t$ to the experiment via the simulation time found above
\begin{align}
    D_{R, exp} = \frac{D}{\Delta t} \,
\end{align}
that, for $\textrm{Pe} = 1000$ and $D = 3.6 \cdot 10^5$, results in $D_{R,exp}~=~14.4~\frac{1}{\text{min}}$, which we can then compare to the inverse of the experimentally measured persistence time, $\tau = \frac{1}{D_{R,exp}}$.

\section{Single-file phase}\label{app:single_file}

During our exploration of the parameter space of the model, we found an interesting cell arrangement that we have not seen previously reported in the literature: Increasing the strength of the repulsive potential in the  differential-CIL setup can produce a phase that has a demixing index of $\ave{M}<0.5$ for high enough packing fractions, see \fref{sfig:singlefile}A. This might seem surprising, since $\ave{M}=0.5$ denotes perfect mixing, so what could be the cause of this ``overmixing''? The answer is the single-file phase shown in \fref{sfig:singlefile}B: At such high densities, the cells approximately form a hexagonal lattice, so that in a single-file configuration each cell will have exactly two same-type neighbours and four opposite-type neighbours, resulting in a demixing index of $\ave{M}= 1/3$, thus driving the overall demixing index below 0.5. 

The cells order themselves into the single-file phase, because it is uniquely stable: At these high packing fractions cells at the edge of a cluster are constantly driving into it, due to the differential CIL mechanism, creating pressure and deforming the cluster and thinning it into a single line. This configuration is stable, because the cells are driving into opposite-type cells which immediately reverse their direction, making the cells bounce back and forth perpendicular to the file, but without pushing up the file. Indeed, we see that the single-file structure disappears as soon as we introduce a delay to the CIL response, because then the cells can push into different-type cells, making the single-file configuration unstable. For weaker potentials, we do not observe this phase, as the repulsive force between same-type cells is not strong enough to keep them lined up.

\section{Author contributions}
AP performed all the experimental work. MB carried out the data analysis, analytical and numerical studies, supported by GP, TB and IB. All authors contributed in the research design, analysis, and interpretation of the results. MB and IB wrote the original draft. All authors revised the manuscript.

\section{Data and code availability}
Data and scripts will be shared upon reasonable request to the corresponding author.

\acknowledgements
IB acknowledges the financial support from Fondo Nacional de Desarrollo Científico y Tecnológico (FONDECYT) project 11230941 and the Universidad de Chile-VID project UI-015/22. EL\ was funded by the President’s PhD Scholarship at Imperial College London.

\bibliography{references}

\onecolumngrid
\pagebreak

\setcounter{section}{0}
\setcounter{equation}{0}
\setcounter{figure}{0}
\setcounter{table}{0}
\setcounter{page}{1}
\makeatletter
\renewcommand{\thesection}{S\arabic{section}}
\renewcommand{\theequation}{S\arabic{equation}}
\renewcommand{\thetable}{S\arabic{table}}
\renewcommand{\thefigure}{S\arabic{figure}}
\renewcommand{\bibnumfmt}[1]{[S#1]}
\renewcommand{\citenumfont}[1]{S#1}

\vspace*{10mm}
\noindent
\textbf{\Large{Supplementary Figures}} \\

\vspace*{3mm}

\begin{figure*}[h]
  \centering
    \includegraphics[width=0.7\textwidth]{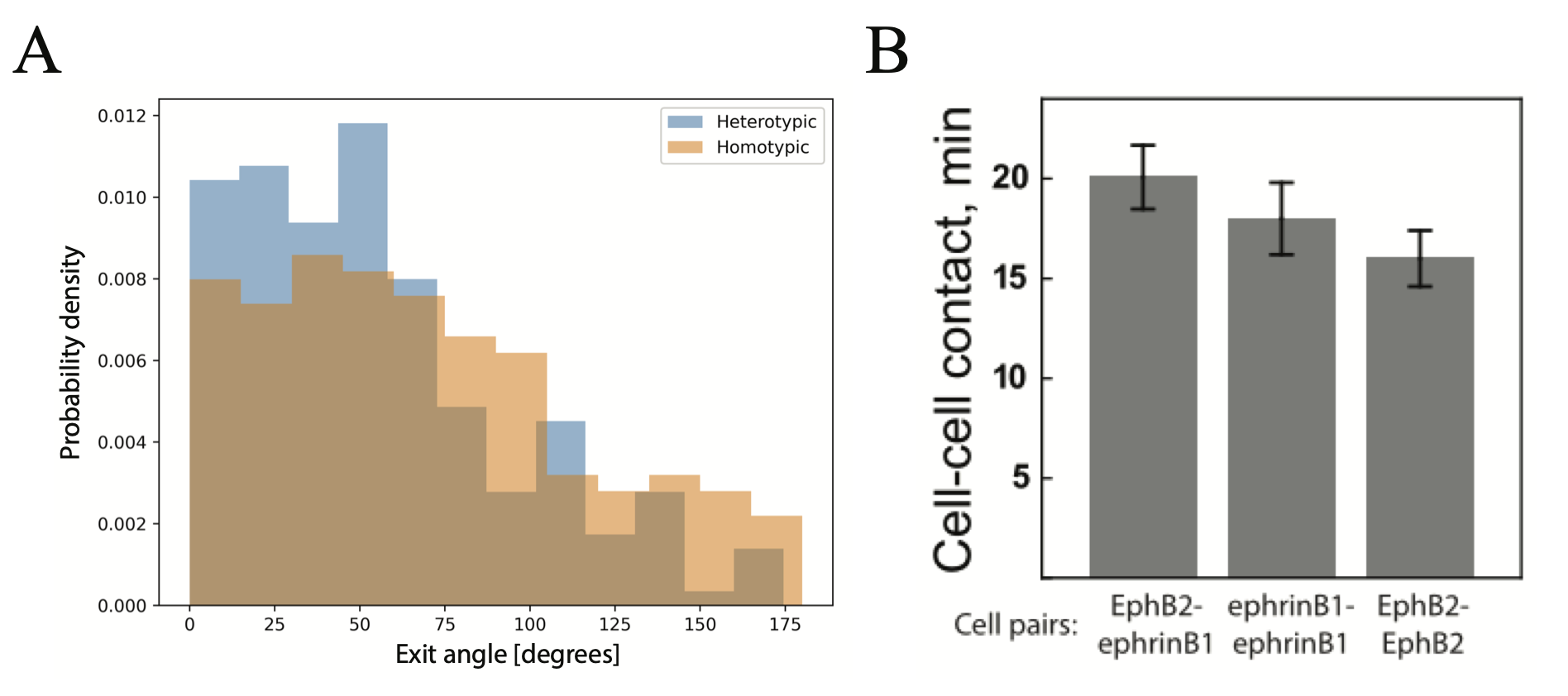} 
  \caption{A) Distribution of absolute values of exit angles for both heterotypic and homotypic contacts. The p-value comparing the two distribution p=0.0003 (Kolomogorov-Smirnov test), indicating a statistically significant but small difference. B) Duration of cell-cell contacts.} 
    \flabel{sfig:stats}
\end{figure*}

\begin{figure*}[h]
  \centering
    \includegraphics[width=1\textwidth]{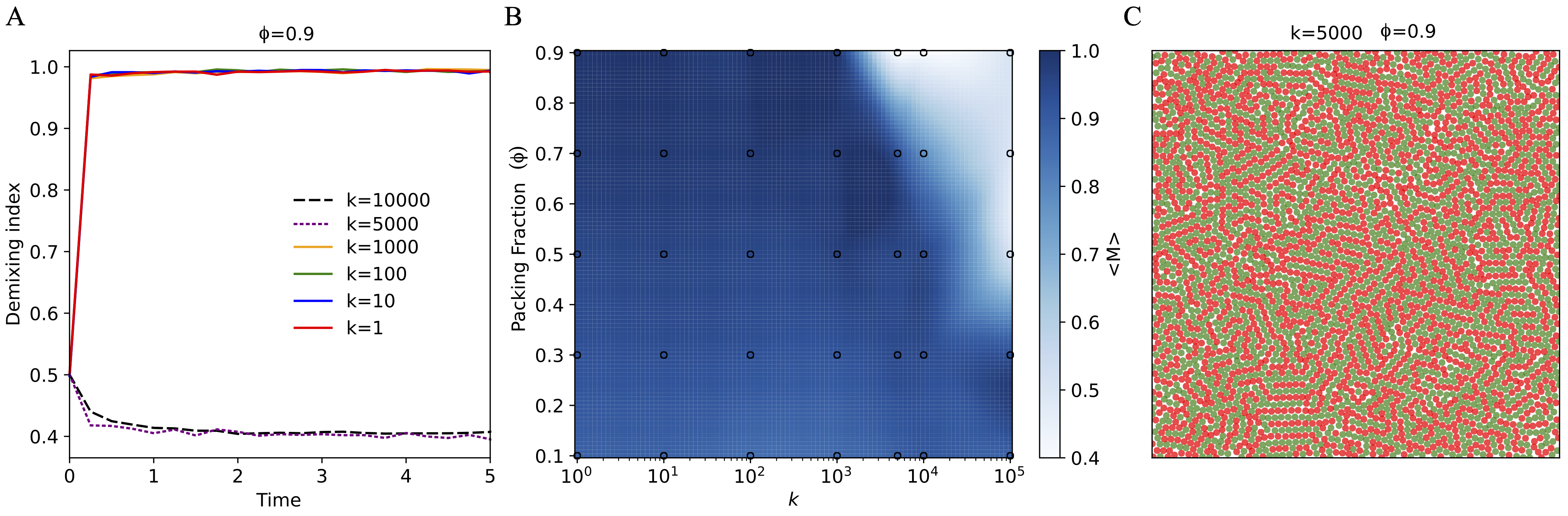} 
  \caption{A) Demixing index as a function of time for heterotypic CIL simulation for packing fraction $\phi = 0.9$ and various strength of the harmonic potential $k$. Above a certain threshold for $k$, the system organises in a single-file configuration, with demixing index below 0.5. B) Demixing index on the $\phi-k$ parameter space for heterotypic CIL simulations. We note that the demixing indices are computed for the configuration at the end of the simulations. C) Shows the single-file state of a realisation at $k=5000$, $\phi=0.9$.} 
    \flabel{sfig:singlefile}
\end{figure*}

\begin{figure*}[h]
  \centering
    \includegraphics[width=0.5\textwidth]{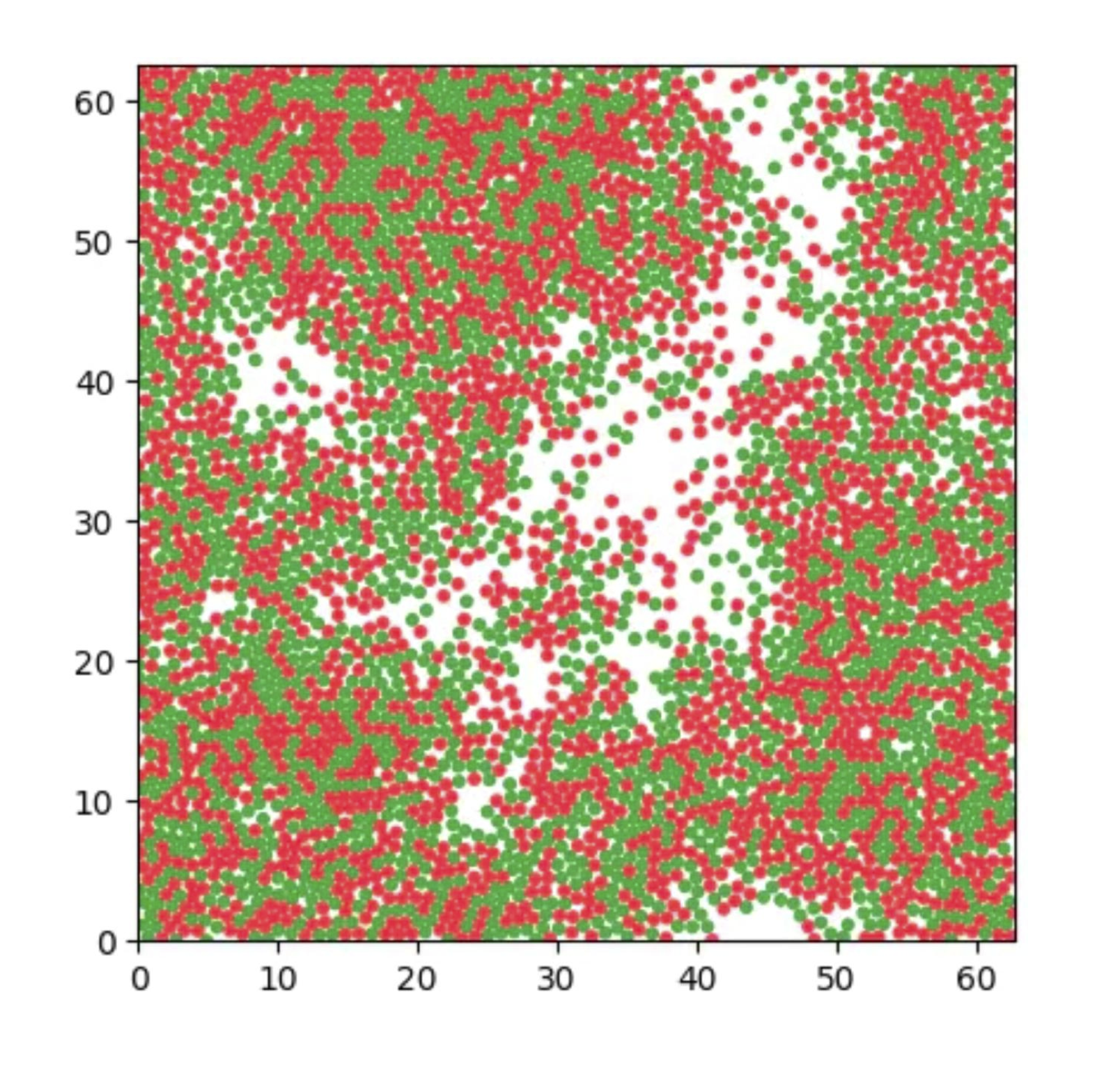} 
  \caption{Result of motility induced phase separation in a simulation where red and green cells are equivalent and there is no sorting mechanisms. 
  The simulation parameters were $\phi=0.8$ and $\textrm{Pe} = 160$.} 
    \flabel{sfig:mips}
\end{figure*}

\end{document}